\documentclass[%
 reprint,
amsmath,amssymb,
aps, %aps
]{revtex4-2}

\usepackage{graphicx}% Include figure files
\usepackage{dcolumn}% Align table columns on decimal point
\usepackage{bm}% bold math
\usepackage{comment}
\usepackage{xcolor}
\begin{document}

\preprint{PRN/123-QED}
\title{
%Perovskite Light Emitting Diodes: Efficiency Roll-off, Sub Band-gap Operation, and Scaling Trends
On the efficiency roll-off in Perovskite Light Emitting Diodes
%Power Dissipation Induced Efficiency and Radiance Roll-off in Perovskite Light Emitting Diodes
%Efficiency Roll-off, Power Dissipation, and  Performance Limits of Perovskite Light Emitting Diodes
%Efficiency Roll-off and Power Dissipation Induced Radiance Limits of Perovskite Light Emitting Diodes
%Efficiency Roll-off and Performance Limits of Perovskite Light Emitting Diodes
%Coupled electrical-thermal-radiative effects and the efficiency roll-off in Perovskite Light Emitting Diodes
%On the near ideal photoluminescence in Perovskites: Is sum of parts better than the whole?
%Near ideal Photoluminescence in Perovskites: Excitonic effects and self consistent back extraction of recombination parameters
%Unraveling the role of excitons in the near ideal performance of perovskite light emitting diodes
}
%Excitonic effects in the photoluminescence of Perovskite Light Emitting Diodes}
%\title{De-mystifying the free carrier - exciton dynamics in perovskite solar cells and light emitting diodes}
% Force line breaks with \\

\author{Pradeep R. Nair}
%\author{et al.}%
 \email{prnair@ee.iitb.ac.in}
\author{Advaith Kiran Marathi}
%\author{et al.}
\affiliation{ Department of Electrical Engineering,
 Indian Institute of Technology Bombay, Mumbai, India\\
}%


%\author{students}
%\affiliation{ Department of Electrical Engineering,
% Indian Institute of Technology Bombay, Mumbai, India\\
%\affiliation{
%\affiliation{
 %Third institution, the second for Charlie Author
%\affiliation{%


\date{\today}% It is always \today, today,

\begin{abstract}
Here, we report a comprehensive modeling framework to unravel the efficiency roll-off in Perovskite light emitting diodes (PeLEDs). Our model self-consistently accounts for a positive feedback mechanism which involves diverse phenomena like temperature-dependent carrier recombination, space charge effects, Joule heating, and thermal transport. Model predictions compare well with experimental results such as dark current-voltage characteristics and efficiency as well as radiance roll-off under high injection conditions. This work identifies key performance limiting phenomena in current state-of-the-art PeLEDs and could be of broad relevance towards device optmization, thermal design, packaging, and operational lifetime.

\end{abstract}

\maketitle

%\tableofcontents

\section{\label{sec:intro}Introduction}
Perovskite based light emitting diodes (LEDs) have achieved impressive progress in the recent years \cite{Yoo2021,Shen2024,Sun2023,Kim2022,longi,Li2024}. Reports indicate external quantum efficiency (EQE) over 30\% \cite{Li2024,kong2024efficient}, and high color purity\cite{Fakharuddin2022}. It is possible to tune emission over a broad spectra by changing the halide composition (and hence the band gap)\cite{dong2020operational}. Due to their excellent electroluminescence properties, Photoluminescence Quantum Yield (PLQY) of halide perovskites has been quite high ($>$ 90\%)\cite{Fakharuddin2022,Li2024}. In addition, high defect tolerance and solution processibility of perovskites contributes to streamline fabrication process \cite{dong2020operational}. Attempts to enhance the performance of PeLEDs include facilitating efficient injection of charge carriers from the transport layers, increasing radiative recombination within the active region, and improving light extraction from the substrate \cite{jia2021excess}. Despite these encouraging improvements, stability remains a major challenge towards successful commercialization \cite{dong2020operational}.\\

For energy-efficient display applications, it is important to achieve high radiance output under low power consumption. Rather, excellent EQE should be demonstrated for large injection conditions (i.e., in terms of the input current). However, initial reports indicate PeLEDs suffer from EQE roll-off (or droop) at high injection levels \cite{jia2021excess,Li2024}. This critical phenomenon is not yet fully understood in terms of the functional dependence of key associated parameters. As a result, there exist several open questions related to the design, optimization, and stability of PeLEDs. To list a few: (a) What causes EQE roll-off in PeLEDs? (c) Is there a consistent explanation for the decrease in radiance at high injection levels? and (c) How does power dissipation and associated Joule heating contribute to efficiency roll-off? and (d) What are the optimization pathways?\\

Here, we provide a comprehensive modeling framework to address the above open questions. Specifically, we unravel the influence of the coupled electrical, thermal, and radiative mechanisms on the operation of PeLEDs. Through this, we predict important features like current ($J$) - Voltage ($V$) characteristics, EQE vs. $J$, energy conversion efficiency (ECE) vs. $J$, radiance vs. power consumption, etc. Our model predictions compare well with recent experimental results. This calibrated model allows us to explore the key mechanisms involved in EQE as well as radiance roll-off in PeLEDs. Curiously, in contrast to the conventional view, here we find that Auger recombination and Joule heating on its own are not the major contributors towards EQE roll-off. On the other hand, a positive feedback mechanism which involves Joule heating and temperature dependence of radiative recombination is the dominant factor that causes efficiency droop in PeLEDs. This help us identify key strategies for further device optimization. Below we first describe a multi-physics model and then compare it against experimental results. \\
\begin {figure*} [ht!]
  \centering
    \includegraphics[width=0.85\textwidth]{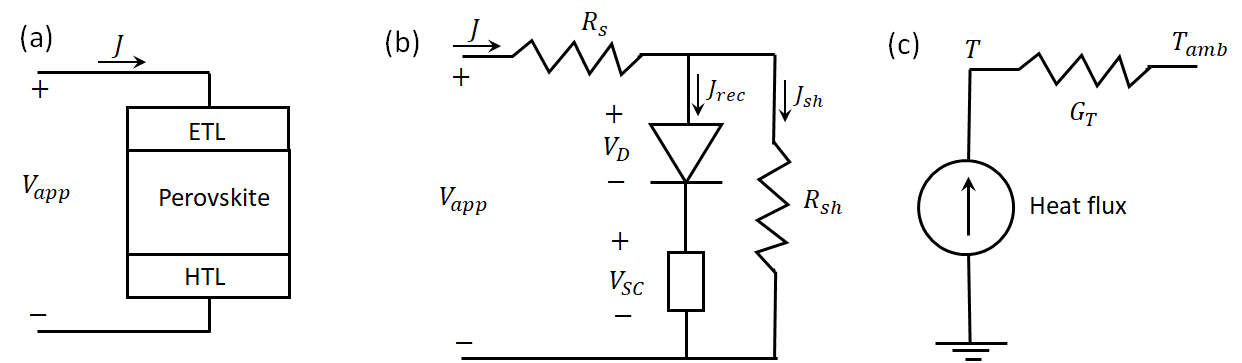}
    \caption{\textit{ Multi-physics model for Perovskite LEDs. (a) Schematic of PeLED. (b) Equivalent circuit description for the electrical characteristics (as described by eqs. \ref{eq:Jrec}-\ref{eq:P}). Here, the applied bias supports the potential drop across $R_s$ and facilitates carrier recombination along with space charge limited transport in the LED. $R_{sh}$ denotes the shunt resistance.  Part (c) shows the equivalent circuit to analyze the heat transport in terms of the power dissipation and effective thermal conductance $G_T$ (as described by eq. \ref{eq:thermal}).}}
\label{schematic}
\end{figure*}

\section{Multi-physics model}
The key performance parameters of a LED are the quantum efficiency, radiance, and the associated power consumption. As such, the main optimization challenge is to achieve the desired radiance with minimum power consumption over the stipulated lifetime. To address this optimization challenge, it is imperative to have predictive models for EQE as a function of injection current. To this end, we need a reasonable description of the JV characteristics and the associated radiative recombination. In this regard, our recent work on photoluminescence (PL) in Perovskites showed that a major factor which contributes to the PL efficiency roll-off (or droop) is the positive feedback mechanism between the laser induced temperature ($T$) increase of the sample and the temperature dependence of radiative recombination \cite{nair2024acs}. Indeed, similar aspects could be relevant for LEDs as well. During operation, the temperature of the active layer could increase due to Joule heating which could influence the efficiency of radiative recombination and hence the EQE roll-off. \\

In this section, we describe a self-consistent model that accounts for the JV characteristics of the device, temperature increase due to Joule heating, and the associated temperature dependence of radiative recombination. For this, we consider a simple PeLED structure with the active layer sandwiched between transport layers (TL, see part (a) of Fig. \ref{schematic}). The carrier transport in PeLEDs can be quite complex and is influenced by phenomena like recombination, space charge effects, ion migration, etc. \cite{Agarwal2014,nandal2017} Although these phenomena are inherently coupled to each other, a compact description of the J-V characteristics is possible through an equivalent circuit representation as given in Fig. \ref{schematic}b.\\

With sufficient electron and hole blocking barriers in the respective transport layers (see Fig. \ref{schematic}a), the current is limited by recombination in the active layer \cite{Agarwal2014,Hossain2024}. Accordingly, the recombination current ($J_{rec}$, see Fig. \ref{schematic}b) is given as 
\begin{equation}
%\begin{align}
     J_{rec} = q(k_1n+k_2n^2+k_3n^3)W_P \\
    \label{eq:Jrec}
\end{equation}
where $q$ is the electronic charge and $W_P$ is the thickness of the Perovskite active layer. This equation assumes that $n=p$, where $n$ and $p$ are the electron and hole densities, respectively. The parameters $k_1$, $k_2$, and $k_3$ account for mono-molecular, bimolecular, and Auger recombination, respectively.  \\

It is well known that recombination processes are temperature ($T$) dependent \cite{PierretADF}. For the temperature range of interest ($300\,\mathrm{K} < T< 350\,\mathrm{K}$), we recently showed that the temperature dependence of radiative recombination can be represented as \cite{nair2024acs}  
\begin{equation}
%\begin{align}
     k_2 = k_{2,F}e^{E_A/kT} \\
    \label{eq:k2T}
\end{equation}
Here $E_A$ is the activation energy, $k_{2,F}$ is the pre-factor, $k$ is the Boltzmann's constant, and $T$ is the temperature of the active layer. Among the several factors, the temperature dependence of refractive indices (or absorption coefficients) significantly contributes to the $E_A$. Using the well known von Roosbroeck-Shockley model \cite{Roosbroeck}, our analysis indicates that $E_A$ could be of the order of $100\,\mathrm{mV}$ - which is supported by transient absorption measurements \cite{Davies2018}. More details on the temperature dependence of $k_2$ is available in our recent publication \cite{nair2024acs}.\\

The applied voltage ($V_{app}$) drops across series resistance ($R_s$, if any) and the PeLED (see Fig. \ref{schematic}b). The potential drop across the PeLED should support both carrier recombination and space charge limited transport. To a first order, this could be visualized in terms of the drop across the active layer ($V_D$, and represented by the diode in Fig. \ref{schematic}b) and the potential needed to support the space charge limited carrier transport ($V_{SC}$). With the assumptions of active layer being undoped,  $n=p$, and negligible drop in the quasi-Fermi levels, we have
\begin{equation}
%\begin{align}
     V_D = \frac{2kT}{q}ln(\frac{n}{n_i}) \\
    \label{eq:vd}
\end{equation}
where $n_i$ is the intrinsic carrier density. Although a single diode is shown in the schematic in Fig. \ref{schematic}b for ease of representation, it is evident from eq. \ref{eq:Jrec} and eq. \ref{eq:vd} that it requires $3$ diodes with distinct ideality factors to represent the recombination current \cite{Hossain2024}.\\

Under steady state conditions, the current through the active layer and the transport layers remains the same. While the magnitude of current is governed by the carrier recombination in the active layer, space charge effects could dictate the carrier transport and hence the net potential drop across the PeLED. Accordingly, in general terms, the potential drop required to support the space charge limited current is given by \cite{duijnstee2020toward,le2021revealing}
\begin{equation}
%\begin{align}
     V_{SC} = K_{SC}J_{rec}^\alpha \\
    \label{eq:vtl}
\end{equation}

where the exponent $\alpha$ could be influenced by the specific nature of trap distribution. We note that a wide range of values (both theoretical and experimental) for $\alpha$ is reported in the literature \cite{duijnstee2021understanding}. Among others, the pre-factor $K_{SC}$ is determined by parameters like mobility, dielectric constant, and effective thickness of the space charge region, etc.\\

Using the equivalent circuit, Fig. \ref{schematic}b, the current voltage characteristics ($J$ vs. $V_{app}$) and the power drawn from the external source ($P$) can now be expressed in terms of the various components as   
%\begin{equation}
\begin{align}
     J &= J_{rec}+(V_D+V_{SC})/R_{sh}
    \label{eq:J}\\
%\end{equation}
%\begin{equation}
%\begin{align}
     V_{app} &= V_D+V_{SC}+JR_s
    \label{eq:Vapp}\\
%\end{equation}
%\begin{equation}
%\begin{align}
     P &= JV_{app}
    \label{eq:P}
\end{align}    
%\end{equation}

The active layer temperature is expected to increase due to Joule heating during LED operation. Of the input power of $JV_{app}$, only a certain fraction is converted to photons which results in light emission. The rest of the input power is dissipated as heat and the thermal conduction through the various layers (which includes packaging) determines the temperature of the active layer. \\

The temperature of the active layer can be effectively estimated using the thermal equivalent circuit shown in Fig. \ref{schematic}c. As the net radiative recombination in the LED is nothing but $k_2n^2W_P$, the corresponding power density of the light emission is given as $k_2n^2W_P\times qE_g \times \eta_{OC}$, where $\eta_{OC}$ is the light out-coupling factor (here we assume that all emitted photons are with energy $E_g$, the band gap). The net power ($P$) delivered by the external source to the LED is $P=JV_{app}$. Accordingly, continuity of the heat flux under steady-state conditions leads us to
\begin{equation}
     G_{T}(T-T_{amb}) = JV_{app} - qk_2n^2W_PE_g \times \eta_{OC} \\
    \label{eq:thermal}
\end{equation}

where $T_{amb}$ is the ambient temperature, and $G_T$ is the effective thermal conductance.\\

Eq. \ref{eq:thermal} describes the balance of heat flux in a PeLED.  Here, the term on the LHS is the net heat flux from the LED to the ambience, while the RHS denotes the net  heat flux generated under electrical injection. Eq. \ref{eq:thermal} assumes that thermal conduction is the main heat transport mechanism, and electrical power dissipation or thermal heat generation happens in the sandwich structure of PeLED (see Fig. \ref{schematic}a). Subsequent heat transport occurs through the various layers (including packaging) to the ambient environment (also see Fig. \ref{schematic}c). Accordingly, the parameter $G_T$ is the effective thermal conductance which depends on several factors such as thickness and thermal conductivity of various layers, interface thermal resistance \cite{organic_TIR}, etc. A similar compact model successfully explained the thermal aspects in thin-film Perovskites under steady state PL measurements \cite{nair2024acs}.\\

The performance parameters of PeLEDs are defined as follows: The internal quantum efficiency (IQE) is defined as the fraction of injected carriers which undergo radiative recombination and is given as
\begin{equation}
     IQE = \frac{k_2n^2W_P}{J/q}\\
    \label{eq:IQE}
\end{equation}

while the external quantum efficiency (EQE) is given as
\begin{equation}
     EQE = IQE\times \eta_{OC}\\
    \label{eq:EQE}
\end{equation} 

The energy conversion efficiency (ECE, or the Wall Plug efficiency) is the fraction of the input power which is emitted as light. With input power given by eq. \ref{eq:P}, and under the assumption that all emission happens at the band-edge, the ECE is given as
\begin{equation}
     ECE = \frac{qk_2n^2E_gW_P \times \eta_{OC}}{J\times V_{app}}\\
    \label{eq:ECE}
\end{equation}
The radiance ($\Phi_{rad}$) in terms of emitted power per unit area per solid amgle under the assumption of single sided emission (with perfect reflection from the other side) is given as 
\begin{equation}
     \Phi_{rad} = k_2n^2W_P\times qE_g\times \eta_{OC}/\theta \\
    \label{eq:radiance}
\end{equation}
where the factor $\theta$ is the effective solid angle to which a small LED can emit photons. \\

\begin {figure*} [ht!]
  \centering
   \includegraphics[width=0.85\textwidth]{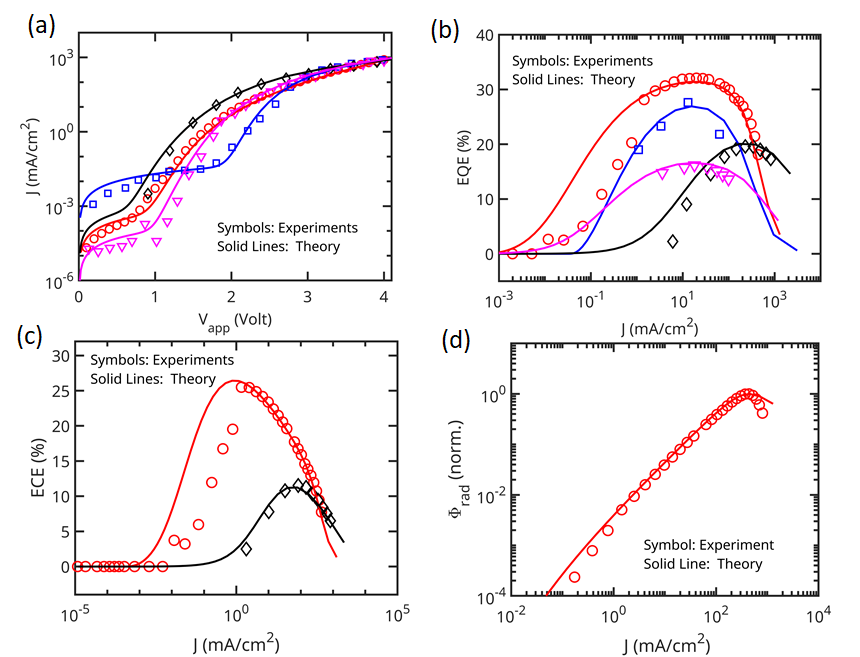}
  \caption{\textit{Comparison of model predictions with experimental results. The symbols represent experimental results from literature while the lines are model predictions (a) Dark $J-V$ characteristics of PeLED, (b) $EQE$, (c) $ECE$, and (d) radiance. The experimental results and corresponding references are: \textcolor{red}{red} circles (ref. \cite{Li2024}), black diamonds (ref. \cite{jia2021excess}), \textcolor{magenta}{magenta} triangles (ref. \cite{zhao2020thermal}), and \textcolor{blue}{blue} squares (ref. \cite{zheng2024ultralow})
  }}
\label{results}
\end{figure*}

\section{Results and Discussions}
Equations \ref{eq:Jrec} -\ref{eq:radiance} provide a self-consistent description of PeLEDs. It accounts for various phenomena like carrier recombination, space charge effects, power dissipation, heat transport, and associated decrease in the radiative recombination efficiency. Numerical solution (Newton-Raphson method) of eqs. \ref{eq:Jrec}-\ref{eq:thermal} leads to self consistent estimates for $n$, $T$, and $J$ for a given $V_{app}$. This allows us to estimate the performance parameters as per eqs. \ref{eq:IQE} - \ref{eq:radiance}. Under large electrical injection, the power dissipation leads to an increase in $T$ (see eq. \ref{eq:thermal}) which leads to a decrease in $k_2$ (as per eq. \ref{eq:k2T}). Any reduction in $k_2$ in turn increases the power dissipation which leads to an increase in temperature (see eq. \ref{eq:thermal}). Indeed, this is a positive feedback mechanism which leads to an increase in temperature that contributes to a reduction in radiative recombination and hence to the efficiency roll-off under high injection conditions. \\

Detailed comparisons between model predictions (i.e., numerical solutions of eqs. \ref{eq:Jrec} -\ref{eq:radiance}) and experimental results are provided in Fig. \ref{results}. The experimental results are from recent publications \cite{Li2024,zhao2020thermal,jia2021excess,zheng2024ultralow}. The experimental results are dark $J-V$ (part (a), 4 data sets), $EQE$ vs. $J$ (part (b), 4 data sets), $ECE$ vs. $J$ (part (c), 2 data sets) and $\Phi_{rad}$ vs. $J$ (part (d), 1 data set). We find several common trends in these experimental results, even though they are from different laboratories, with different active layers of perovskites ($E_g$ ranging from $1.4-2.42\,\mathrm{eV}$) and different processing conditions. To list a few: (i) low bias regime of $J$ vs. $V_{app}$ shows influence of shunt resistance, (ii) diode like features in $J$ vs. $V_{app}$ in the medium bias regime, (iii) nearly similar sub-exponential increase in $J$ for large $V_{app}$ (Fig. \ref{results}a), (iv) a broad peak in both EQE vs. $J$ and ECE vs. $J$ followed by roll-off, and (v) a near linear increase in radiance with $J$ followed by its roll-off under high injection conditions (Fig. \ref{results}d). \\ 

Model predictions are shown as solid lines in Fig. 2. Indeed, all the key trends in experimental results are well anticipated by our proposed model, which validates the multi-physics approach. The parameters used in this study are listed in Table \ref{tab:my_label}.  The parameters used in the model predictions for the experimental results from ref. \cite{Li2024} (i.e., data set shown in red color, Fig. \ref{results}) are based on a self-consistent analysis of the corresponding time resolved Photoluminescence (TRPL) transients (as recently reported \cite{nair_exciton}). We find that model predictions indicate that the parameters related to recombination kinetics ($k_1, k_2, k_3$), light out-coupling factor ($\eta_{OC}$), space charge limited transport ($k_{SC}, \alpha$) etc. are comparable in their relative magnitudes, with some key distinct trends,  across the different experimental data sets.\\

Several noteworthy insights are readily available in the results shared in Fig. \ref{results}, as discussed below. \\ 

\begin{table*}[t]
    \centering
\begin{tabular}{|p{15mm}|p{20mm}|p{28mm}|p{28mm}|p{30mm}|p{28mm}|}
\hline
Parameter &Units&\multicolumn{4}{c|}{Fig. \ref{results}: Data sets}\\
\cline{3-6}
%&&&&&\\
%\hline
%&&&&&\\
%\cline{1-6}
%&&&&&\\
%\hline
       &  & Model: \textcolor{red}{red} line, Exp: \textcolor{red}{red} circles, ref. \cite{Li2024} & Model: black line, Exp: black diamonds, ref. \cite{jia2021excess}  & Model: \textcolor{magenta}{magenta} line, Exp: \textcolor{magenta}{magenta} triangles, ref. \cite{zhao2020thermal} & Model: \textcolor{blue}{blue} line, Exp: \textcolor{blue}{blue} squares, ref. \cite{zheng2024ultralow}\\
        \hline
%         Data set in Fig. 2& & red circle & black diamond & magenta triangle & blue square \\
%        \hline

         \hline

         \hline    

         \hline

         \hline

         \hline      
  
         \hline
 
      \hline

         \hline

         \hline

         \hline

         \hline

         \hline

         \hline
         
    \end{tabular}
    \caption{Parameters used in model predictions in Fig. 2. Here, $E_g$ and $W_P$ for each data set are as reported in the respective references. The recombination parameters used in the analysis of experimental results from ref. \cite{Li2024} are based on an analysis of corresponding experimental TRPL data (ref. \cite{nair2024_exciton}) } 
    \label{tab:my_label}
\end{table*}

\textbf{A. Efficiency roll-off and Joule heating}: An important tradeoff associated with LEDs is the desired light output and associated power consumption. The same is usually discussed in terms of efficiency roll-off under high injection conditions. The results shown in Figs. \ref{results}b,c show that the proposed model compares well with the $EQE$ as well as $ECE$ characteristics of experimental results. For LEDs limited by carrier recombination, the low bias current could be dominated by monomolecular recombination (with $J\propto k_1n$) while bimolecular recombination dominates the mid- bias regime (with $J\propto k_2n^2$). Under large biases, the current could be dominated by Auger recombination (with $J\propto k_3n^3$). Accordingly, it can be shown that the $EQE$ first increases linearly with $n$, then it may remain fairly independent of $n$ and finally the $EQE$ will vary as $n^{-1}$. In  terms of $J$, we expect the $EQE$ to first increase with $J$, then remain invariant with $J$, and finally decrease as $J^{-1/3}$. Hence, $EQE$ roll-off is a fundamental characteristic expected in all LEDs where the current is limited by carrier recombination.\\

\begin {figure} [h!]
  \centering
    \includegraphics[width=0.45\textwidth]{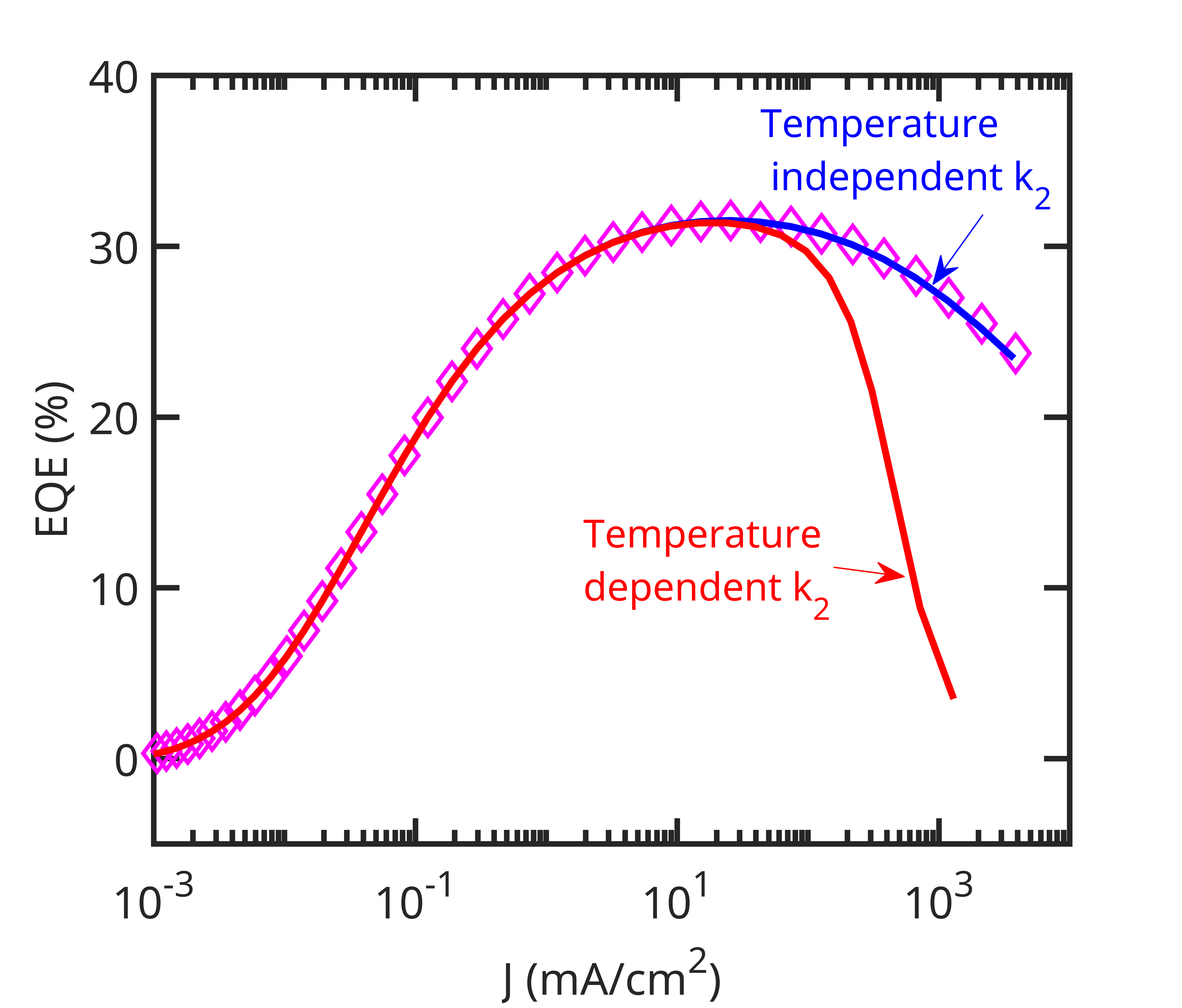}
    \caption{\textit{Influence of Joule heating on $EQE$ roll-off. The symbols represent simulation results under room temperature operation (i.e., $T=298K$). Here the roll-off is entirely due to Auger recombination. Blue line indicates $EQE$ roll-off in the presence of Joule heating with $k_2$ being temperature independent. The red line indicate $EQE$ roll-off in the presence of Joule heating with temperature dependent $k_2$. These results indicate that Joule heating on its own do not have any significant contribution to $EQE$ roll-off. In fact, $EQE$ roll-off is dominated by the Joule heating induced reduction in $k_2$. The parameters used are the same as the data set indicated by red lines and symbols in Fig. \ref{results}.
     }}
\label{Temp_effect.png}
\end{figure}

The $EQE$ roll-off in PeLEDs is often attributed to Joule heating effects - in addition to the above discussed influence of Auger recombination. Our model provides interesting quantitative insights in this regard. Figure \ref{Temp_effect.png} compares the influence of Joule heating and temperature dependence of $k_2$ on the $EQE$ roll-off. Here, the symbols represent simulation results at room temperature - i.e., in the absence of any Joule heating effects with $k_2$ being a constant. The $EQE$ roll-off in this case is entirely due to Auger recombination being the dominant mechanism under high injection conditions. The blue line shows simulation results in the presence of Joule heating but with $k_2$ being temperature independent. Surprisingly, we find that the $EQE$ roll-off remains unchanged in this case. This indicates that Joule heating, on its own, does not contribute to any additional roll-off in $EQE$. This is expected as $EQE$ in the absence of shunt paths (see eq. \ref{eq:EQE}) depends only on the recombination parameters and $\eta_{OC}$. With these parameters being temperature independent, Joule heating is not expected to contribute to $EQE$ roll-off. On the other hand, with $k_2$ being temperature dependent, we find that $EQE$ roll-off is significantly affected (see red line in Fig. \ref{Temp_effect.png} which is also the same data plotted in Fig. \ref{results}b). Hence,  $EQE$ roll-off in PeLEDs is caused by a combination of Joule heating, the temperature dependence of $k_2$, and increased Auger recombination (and not by a stand alone effect or a combined effect of Joule heating and Auger recombination). \\

\textbf{B. Radiance roll-off and leakage currents}: Our model helps to delineate the role of parasitic elements, such as leakage paths, in the PeLED characteristics. The influence of simple shunt resistances is often directly evident from the $J-V$ characteristics (see Fig. \ref{results}a). However, it is possible to have several parasitic effects which are more complex than simple shunt resistances. For example, the energy band offsets or barriers between the transport layers and the active layer might not be large enough. Here, the carriers injected from one TL could reach the other TL without undergoing recombination. Such undesired over-the-barrier transport of carriers also contributes to $J$ and hence to the $EQE$ roll-off under high injection conditions. Can the proposed formalism be used to ascertain and quantify the influence of such non-ohmic parasitic components? Rather, what signature could unambiguously distinguish between the power dissipation-induced temperature increase and its after effects against parasitic leakage paths?\\ 

The key to unravel the above puzzle lies in the scaling trends of radiance (and not the $EQE$ roll-off). As $J$ increases, the radiance of a LED, which depends only on the radiative recombination, is expected to increase monotonically (it may saturate, in some limiting scenarios). This trend is expected even in the presence of additional undesired over-the-barrier transport of carriers. The radiance can decrease only if there is a decrease in $n$ or $k_2$ or both. Typically, any decrease in $n$ will also lead to a decrease in $J$ (both recombination and parasitic components). Hence, an experimental observation of a decrease in radiance with an increase in $J$ can happen only if $k_2$ decreases under high injection conditions (unless dominant non-radiative  recombination pathways are newly activated through material degradation or otherwise). The same is predicted by our model and is seen in experiments as well (see Fig. \ref{results}d). Hence, instead of the $EQE$, the radiance vs. $J$ trends could provide unambiguous evidence on the underlying physical mechanism that govern the PeLED operation under high injection conditions - leakage paths vs. a positive feedback involving power dissipation, heat transport, and reduced radiative recombination.\\

\textbf{C. Optimization pathways}: Our results identify an important performance limiting bottleneck in some of the recent PeLEDs. As indicated by Fig. \ref{results} and Table \ref{tab:my_label}, the ON characteristics is dominated by space charge limited transport. This increases the operating voltage $V_{app}$ required to achieve significant radiance. The corresponding power dissipation increases temperature which contributes to an reduction in $k_2$ and hence the $EQE$ roll-off. Hence, the most important optimization pathway for PeLEDs is to improve the transport properties - both electronic and thermal. This will aid to achieve high radiance with reduced power consumption. Of course, improvement in material quality (i.e., reduced $k_1$) is a also a desired feature. \\

The increase in active layer temperature and the associated decrease in $k_2$ are central to $EQE$ roll-off. This reduction in $k_2$ could be due to either $E_A$ or $T$ or both (see eq. \ref{eq:k2T}). Eq. \ref{eq:thermal} indicates that $T$ is significantly influenced by $G_T$. Accordingly, the variables that dictate the $EQE$ roll-off are $E_A$ and $G_T$. Larger $E_A$ or smaller $G_T$ leads to a significant reduction in $k_2$ and hence increases the $EQE$ roll-off. This indicates the importance of characterizing the temperature dependence of the recombination parameters and appropriate thermal design for PeLEDs.\\

\textbf{D. Dark JV Characteristics}: Figure \ref{results}a indicate that the dark JV characteristics share some common features. For example, the low bias regime of all devices show evidence of shunt resistances over a broad range of values (i.e., over two orders of magnitude, see Table \ref{tab:my_label}). The turn-ON characteristics (i.e., with diode-like exponential increase in current with $V_{app}$) shows significant variation. However, the same is consistent with the variation of $n_i$ with $E_g$. As the $E_g$ increases, $n_i$ decreases exponentially and hence the $V_{app}$ required to turn the device ON increases (as per eq. \ref{eq:vd}). The back extracted values for $n_i$ in Table \ref{tab:my_label} varies almost exponentially with $E_g$ (the deviations could be attributed to the differences in the effective density of states of the respective materials). The high bias regime of all devices shows similar trends; however, these are not limited by series resistance effects. On the other hand, these characteristics are well described by the space charge effects (i.e., eq. \ref{eq:vtl}) and the corresponding parameters ($K_{SC}$ and $\alpha$) are similar for all devices -  a surprising result as these devices have different materials, structures, processing conditions, etc. We find $\alpha=0.25$, which is consistent with the literature \cite{duijnstee2021understanding,bai2023perovskite}. Future research could explore device architectures to reduce the operating voltages and hence power consumption. \\ 

Overall, here we developed a coherent theoretical framework for PeLEDs which accounts for several critical phenomena like temperature dependent carrier recombination, space charge limited carrier transport, thermal dissipation, heat transport, and a positive feedback mechanism which involves all of the above. As a result, our model anticipates several key features of PeLEDs (with band gaps ranging from $1.4-2.42\,\mathrm{eV}$), and elucidates the scope and pathways for further optimization. The model predictions are indeed subject to the parameters used. Here, the recombination parameters associated with one data set (shown in blue, Fig. \ref{results}) were back extracted from the corresponding TRPL data (see ref. \cite{nair2024_exciton}). Of the various parameters listed in Table \ref{tab:my_label}, $k_1, k_2,$ and $k_3$ dictate the carrier recombination while $K_{SC}, \alpha, R_{s}, R_{sh}$, etc. influences the transport characteristics (i.e., large bias dark $J-V$). Accordingly, our simulations indicate that small changes in $\alpha$ lead to distinct changes in the $J-V$ characteristics (with insignificant changes in $EQE$ vs. $J$ characteristics, see Fig. S1, Suppl. Mat.), while $EQE$ and $ECE$ are very sensitive to even minor changes in recombination parameters (with negligible changes in $J-V$ characteristics, see Fig. S2, Suppl. Mat.). As such, this highlights a unique aspect about the proposed model: We describe several features (i.e. $JV$, $EQE$, $ECE$, radiance, etc.) with contrasting dependence on key functional parameters through a coherent modeling framework which establishes the validity of both the methodology and the parameters used.  Additional details on the variation of $V_D$, $n$, and $T$ as a function of $V_{app}$ are provided in Fig. S3, Suppl. Mat. We also note that the parameters $T$ and $E_A$ have a combined effect on the device characteristics. Similar amount of $EQE$ roll-off can be obtained for a lower increase in $T$, but with a larger $E_A$ and vice-versa.\\

It is well known that PeLEDs (and solar cells) could be influenced by aspects like ion migration, band level offsets, interface recombination, etc. The influence of the same on solar cells has been addressed in several of our prior publications \cite{nair_acsEL,sumanshu_apl,nandal2017,abhimanyu_JAP2021}. In particular, the assumption $n=p$ in the active layer might not be appropriate under all scenarios \cite{jia2021excess}. However, the same holds in the presence of significant mobile ions and hence the insights shared are relevant for such cases. We note that our model overestimates EQE and ECE for low J values (see Fig. \ref{results}b,c). This is not a major concern as such low current levels (i.e., $J< 1\, \mathrm{mA/cm^2}$) are not of practical significance. A possible physical origin for this mismatch is explored through numerical simulations in Suppl. Mat. (dependence on mobile ions, see Fig. S4). Further, experimental results indicate that the radiance roll-off at very high injection conditions is more than the model predictions (see Fig. \ref{results}d). Several non-idealities including material degradation and hence an increase in non-radiative recombination could be relevant at such extreme high injection conditions.  In addition, the power dissipation and associated thermal analysis provided in this manuscript could be relevant for modeling the operational lifetime of PeLEDs in view of various degradation mechanisms such as phase segregation\cite{abhimanyu_JAP2021}, trap generation, grain boundary effects, etc. Our results also have obvious implications towards back extraction of recombination parameters from $EQE$ characteristics of PeLEDs. \\

\section{Conclusions}
In summary, here we report a comprehensive modeling methodology to identify key performance limiting factors of PeLEDs. Our analysis shows that undesired power dissipation due to space charge effects and associated thermal management issues leads to a temperature increase in the active layer. In addition, a positive feedback mechanism between heat transport and temperature dependence of radiative recombination results in several important features, such as efficiency and radiance roll-off under high injection conditions. As such, this manuscript provides a multi-physics description of PeLEDs with key performance limiting factors identified, paving the way for further optimization. 

\section{Acknowledgements}
The authors acknowledge National Center for Photovoltaics Research and Education (NCPRE), Indian Institute of Technology Bombay. The authors also acknowledge discussions with Simhadri Venkata Ramana, and Sushama Usurupatti, IIT Bombay\\

\section*{References}
\bibliography{apssamp}% Produces the bibliography via BibTeX.

\end{document}